# On Web-based Domain-Specific Language for Internet of Things

Manfred Sneps-Sneppe
Ventspils International Radioastronomy Centre
Ventspils University College
Ventspils, Latvia
manfreds.sneps@gmail.com

Dmitry Namiot
Faculty of Computational Mathematics and Cybernetics
Lomonosov Moscow State University
Moscow, Russia
dnamiot@gmail.com

*Abstract*—This paper discusses the challenges of the Internet of Things programming. Sensing and data gathering from the various sources are often the key elements of applications for Smart Cities. So, the effective programming models for them are very important. In this article, we discuss system software models and solutions, rather than network related aspects. In our paper, we present the web-based domain-specific language for Internet of Things applications. Our goal is to present the modern models for data processing in Internet of Things and Smart Cities applications. In our view, the use of this kind of tools should seriously reduce the time to develop new applications.

I. INTRODUCTION

In this paper, we would like to discuss the software models and architectures for the Internet of Things (IoT) programming. There are many papers devoted to the network related aspects of Machine-to-Machine (M2M) and IoT [1]. Our goal in this paper is to discuss system software models and solutions. This paper continues our series of publications about software aspects of M2M and IoT.

The term IoT was introduced in a paper [2] as a union of Internet-connected sensors, devices, and citizens. There are many definitions for the term Smart City. For examples, the authors in [3] describe it as an Internet-connected web of citizens (people) and electronic sensors/devices (things) that can serve many functions related to public and environmental health surveillance and crisis management applications. The key moments here are human related aspects. Sao, we can conclude, that in many aspects IoT is an engine for Smart Cities applications. And M2M domain is shortly the IoT domain without user interfaces (UI). Obviously, the UI is a mandatory part of IoT projects and could be missed in M2M. This definition lets us make a more radical statement: M2M is simply a part for IoT. For our programmers-oriented (data access oriented) approach this definition is, probably, most suitable.

The effective model for IoT application is a hot area attracts a big attention. There are many papers discusses the ways for common software standards in IoT area [4]. It is especially important due to the high diversity of sensors and devices.

In paper [5] authors discussed the problems with the unified standards of Machine to Machine communications (M2M). They concluded that the current development misses the larger point of how M2M services and products get created and deployed. In many cases, developers either have to use some predefined platform and be locked with its restriction or build a system completely from scratch. For M2M and IoT products to be successful, interfaces (programming interfaces) must be simple. The complexity that lies underneath should be completely hidden from the developers. As seems to us, at the current stage the existing and proposed solutions very often just increase the complexity.

The complexity of existing approaches is also discussed in paper [6]. It raises the following question: do we really need Application Program Interfaces (API) always, or our goal could be described as Data Program Interfaces (DPI)? We can describe DPI as an interface at the edge of an IoT device that exposes and consumes data. IoT devices very often do not support commands (instructions). Many of sensors just provide some data and nothing more. This simple step (refusal to support API) can seriously simplify the interaction with the devices. DPI's are much simpler, of course. And what is more important – they can create a unified API for all devices. The process of reading data can be similar for all devices. As usual, we can pass data interpretation (translation) to the end-user devices. And our "unified" reading procedure can simply return some JSON array.

So, as soon as all the "unified" standards become too complex, what is the solution? We are strong proponents of micro-services.

The micro-services approach is a relatively new term in software architecture patterns. The micro-service architecture is an approach to developing an application as a set of small independent services [7]. Each of the services is running in its own independent process. Services can communicate with some lightweight mechanisms (usually it is something around HTTP) [8]. Such services could be deployed absolutely independently. Also, the centralized management of these services is a completely separate service too. It may be written in different programming languages, use own data models, etc. We think that micro-services are the natural fit for M2M (IoT) development.

In accordance with this, in our opinion, considering the individual systems, such as Open IoT [9], for example, a description of their abilities cannot be the main purpose. The main point is the allocation of micro-services within them. And the second goal is, accordingly, the issues of their independent usage and deployment. Such an analysis with respect to M2M applications was presented in our paper [10].

IoT and M2M have remote device access in common. But they are not completely similar, of course. Some of the authors draw the difference in the way IoT and M2M access to the remote devices. For example, traditional M2M solutions typically rely on point-to-point communications using embedded hardware modules and either cellular or wired networks. In contrast, IoT solutions rely on IP-based networks to interface device data to a cloud or middleware platform [11]. It is probably now always true because the cloud is not a mandatory stuff for the Internet of Things. We think that this statement is very important. Nothing prevents the application access to remote devices directly, or, more precisely, get data from remote devices without the cloud (and without the middleware, by the way). The typical examples are Bluetooth Low Energy tags, mentioned in [6]. The network related aspects (protocols) are out of the scope of this paper. So, we are not going to discuss IP vs. non-IP networks.

The rest of the paper is organized as follows. In Section II, we discuss the common challenges of IoT programming. In Section III, we discuss perspective programming models and software architecture approaches for IoT applications.

## II. IoT PROGRAMMING CHALLENGES

As the first challenge for the system development in IoT area, we should mention the power supply. Obviously, it is the first limitation. It directly affects the algorithms we can use in our systems. So, solutions (e.g., libraries) for implementing power-optimized calculations (algorithms) will prevail. The same is true for network protocols.

We should mention in this context such entity as Dynamic Power Management (DPM). The main idea behind this approach is to shut down devices when they don't need to be on-line on and to start them up when they need to transmit (receive) data. As per [12], Dynamic power management (DPM) is a design methodology for dynamically reconfiguring systems to provide the requested services and performance levels with a minimum number of active components or a minimum load on such components.

Normally, it is a typical task for the operating system (OS). E.g., a mobile operating system can prefer accelerometer over GPS for some tasks due to energy limitations, etc. But complex IoT may orchestrate several devices, and any individual operating system is simply unaware about the whole process. So, the whole system should be able to switch services on and off more intelligently than each individual device's OS.

But, of course, DPM itself is not free and may cause such a problem as latency. The latency could be of course a congenital problem for IoT devices too. E.g., a device may transmit data in discrete time cycles only. The typical example is the above-mentioned BLE tag (iBeacon).

Another typical source of delays is very often the network topology optimized for IoT system. For example, mesh networks are immune to the failure of a few nodes [13]. But at a price for this we will have more hops (read – increased delay) in data delivery paths. Actually, the scalability for IoT networks is a big problem. The things could be more complicated if we admit the fact that many devices may simply transmit data without requests (e.g., perform some operations by the own timer or due to some external activity). It could lead to the wasted bandwidth and increased delays in communications.

The reliability is the next big issue. The whole set of devices could be constantly checked, for example. So, in general, for many use cases we have to consider IoT data as unreliable. It may lead to the additional data curation and error-correction procedures on the application level [14].

The data curation and data brokering stuff is very important for IoT applications by the another reason also. Remote devices (sensors) in case of IoT can produce a huge amount of data. And it is very important to have the ability for data projection. We need to select the right amount of data for the particular task. And one of the biggest problem here is to find a right (and commonly used) tool just for data description. Raw data from sensors should have some meta-data associated with them. Otherwise, there is no way to develop the adaptive algorithm. As soon as the mapping for data is unknown, we cannot automatically detect the dependencies for example. And this information is critical for many algorithms.

Figure 1 illustrates the basic data model behind FI-WARE project [15].

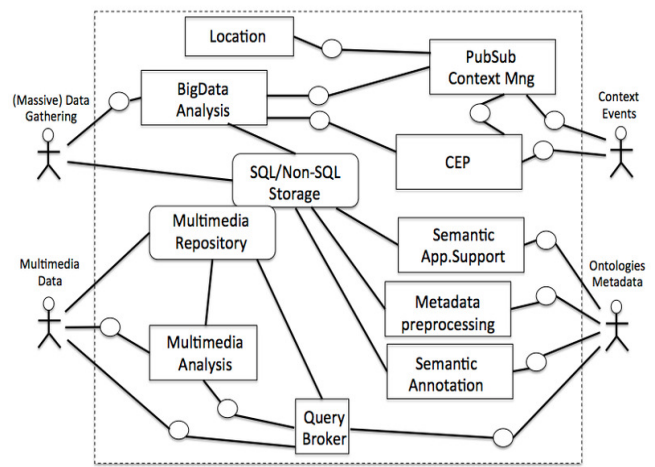

Figure 1.    FI-WARE model

Obviously, remote devices (sensors) may generate a big amount of data. So, the Big Data approach is a natural fit for IoT. But in case of a huge amount of distributed data developers need a way for real-time processing some sub-sets. Think, for example about processing sensors data for some limited retail space. It means, there is a huge demand for some kind of toolchains [16].

Current IoT architectures are devices or networks-oriented. However, the key value proposition of IoT is from the interaction of these "Things" with humans and society. So, for getting the benefits, some form of stream processing for IoT data is practically mandatory.

It the terms of context-aware computing ("ubiquitous computing"), IoT makes the software context much larger. So, the developed applications should have some mechanisms for dealing with this fast changed data.

### III. IoT PROGRAMMING MODELS

For many years academic papers discussed web services models for IoT (M2M, Smart Cities) [17]. We may discuss underlying protocols (MQTT, COAP, etc.) [18], data formats (XML vs. JSON), but this does not negate the fact of a very low granularity for web requests. Yes, the web services can unify the programming (coding practices), but due to the huge amount of the various devices in the programming level, we have to deal with a plenty of asynchronous requests and manage them inside of the code manually.

The industrial approach to programming requires the componentization. The code should deal with components. In general, any component should contain the program interface (API or DPI). But it is not enough for IoT or M2M. APIs for components (e.g. for sensors) exist right now. For IoT, we should include the behavior (the calculations) into components too. It is the way to radically simplify the development. Simply, the developer should be able to obtain alarms, history, predictions right from some block of sensor, rather than pull raw measurements data and perform all the above-mentioned functionality again and again in the own code. In other words, accepting the convenient unification from Web technologies, we need to redefine the word "resource". What could be behind universal resource identifier (URI) in the modern web? Let us see some models in this connection.

In the paper [19], the author introduces CREST (Computational REST) model and provides the definition of a resource as a locus of computation. As per RFC 2396 [20], a resource can be anything. Looking at the definition of a resource, we can distinguish between three elements: a resource; the state of a resource; and the representational state. By definition, resources can never be accessed and are only manipulated through their representations [21]. CREST is based on the following axioms:

- A resource is a locus of computations, named by an URL.

- The representation of a computation is an expression plus metadata to describe the expression.

- All computations are context-free.

- Only a few primitive operations are always available, but additional per-resource and per-computation operations are also encouraged.

- The presence of intermediaries is promoted.

Let us see on the basis of existing models. Web of Things is the most typical example for our explanation [22]. The fact that we can use HTTP requests (e.g., REST as XML or JSON over HTTP) to obtain information from a single device (sensor) does not help when a large (huge) amount of devices is deployed. Yes - it makes programming a much more uniform, but for a large set of devices, the developer must still manually organize polls for devices in order to collect data, organize wait cycles and synchronization, etc. and forward the results of the synchronization.

We must always deal with the fact that most our devices (sensors) are asynchronous. As an analogy here we can mention MapReduce [23] approach. Technically (from the development point of view) it is just a library (Java package), which helps to organize the parallel execution of threads and assemble the results.

By this reason, we believe that the IoT programming will require a paradigm shift. A simple declaration that we will use XML over HTTP is not enough. We should think about some tools for closing the gap between the distributed platform and sequential computing paradigm (sequential programming languages and frameworks).

In our paper [24], we analyzed the typical IoT applications for wireless tags (iBeacons). We can present the top-level definition like this:

- There is a set of sensors we need to poll periodically for getting new measurements

- There is a set of sensors we need to accept data from (push data – sensors initiated communications)

- The business process could be presented as a set of productions (rules). Each of the rules depends on some available data and, probably, on some global variables (states).

- The data availability always assumes the presence of data for any finite set of timestamps. In other words, the application makes conclusions (actions) depending on some window of measurements.

The last statement fixes the fact that in the most cases IoT application deals with the finite set of the "latest" measurements. So, for example, our processing will deal with the latest measurement (timestamp $t$) and some recent history ($t-1, t-2,$ etc.) Of, course, for some tasks (e.g., billing as the most obvious example) we will need the whole timeline, but we are talking about the majority of applications.

The next important moment is the meaning of the word "availability". We assume that data are available, when the application receives a chunk of data, the process of receiving is completed, and data are available for processing.

We see the future IoT programming in the declarative models. Declarative networking [25] is an approach that promotes declarative, data-driven programming to concisely specify and implement distributed protocols and services. Datalog [26] is, probably, the most known example of rule-based language in this area.

Classically, declarative programming is a programming paradigm that expresses the logic of a computation without describing its control flow. In declarative programming, we have to specify what is to be computed, rather than how it is to be computed. The idea is to avoid a detailed description of our algorithm of computation. We should leave this part (algorithm) the some automatically generated applications. By this reason, all declarative programming systems contain at least two components: a programming language and its execution system. The classical example is well known SQL. Writing database queries in SQL could be considered as declarative programming. At the same time, by the practical reasons, many programming languages are hybridized and contain both declarative and imperative language constructs. The good example is any extension of SQL. Or even some of the SQL operators: SELECT is declarative, UPDATE and DELETE are imperative.

Originally, a Datalog program consists of a set of deductive rules. Each rule has a rule condition (head) and a rule body, which are separated by the deduction symbol *:-*. A condition is a relation name with variables or constants as arguments. A rule's body is a list of predicates or Boolean expressions that conjunctively derive the predicate in the corresponding rule condition. A relation can be defined as either in an extensional database or an intensional database. EDB relations present the inputs of the program. They should remain constant during the life time. IDB relations are deduced based on EDB (and other IDB) relations, according to the rules. The typical form is:

```
r1 rule1(A,B) :- action1(A,B).
r2 rule2(C,D) :- action2(C,D), rule1(C,D).
```

Here the second rule refers to the first one. These rules are declarative because they only specify what conditions the reachable relation should satisfy, rather than procedures to compute it.

But in the same time, on practice, each implementation is de-facto some hybrid systems. For example, even for the pure declarative system we need some procedural things for creating a user interface, for printing reports, etc.

Our prototype LogicIoT allows programmers to write applications without being concerned with low-level programming details. The prototype is implemented as an extension for JSP pages as a set of custom JSP tags.

The language model contains the following objects: relations, triggers, endpoints, timers, facts, rules and modules.

Relations are analogues for relations (or data tables in relational databases. E.g.:

```
RELATION R (MAC, RSSI)
```

The above-mentioned relation describes data tuples for wireless networks. Each record has got a field *MAC* (we will keep here a MAC-address) and *RSSI* (signal strength)

Each record has got automatically added timestamp field. Its name is *T*. Here we follow to the standard practice of NoSQL models (e.g., time stamped records in Cassandra [27]).

The fact that we use the word 'relation' and analogues with relational databases is not associated with an automatic representation of data as relational tables. For each relationship, there will be specified a module for its implementation, which will hold the data view. The implementation module is the code (in our case - JSP) file that implements the operations of adding data, read data, as well as an analogue for trigger INSERT operation in relational databases. Since in our case the implementation of the module is a JSP file, the support of these operations is as follows. To add data operations, JSP file handles a separate HTTP GET request, where the parameters describe the data for new records. For a read operation, JSP file supports a separate HTTP GET request that returns JSON array with the requested record. For the implementation of the trigger, JSP module refers to some given (predetermined) URL, passing fields for a fresh record in the parameters. In other words, a trigger performs HTTP GET request for the each new record added to our relation.

If R is a name for our relation, then *R.MAC* describes a value for the field named MAC in the latest (by the time) record. *R.MAC*[-1] describes a value for the field named MAC in the last but one record and so on.

The form *R* (value1, value2) means adding a new record to the relation *R* with the given values.

For the each relation, we can optionally define a trigger. The trigger is a block that our system will execute as soon as a new record is added to the relation. So, any existing trigger is always associated with some relation. It is a block of code executed during adding a new record to the given relation. The closest analogue is TRIGGER INSERT in relational databases. If our code adds multiple entries, then the trigger is executed multiple times (an analogue is TRIGGER INSERT FOR EACH ROW).

```
TRIGGER (R)
{
}
```

The trigger in LogicIoT always defines post-processing actions. In other words, it will be executed after adding a new record. So, in the above-mentioned example, we can use an expression like *R.RSSI* in our trigger and this expression will define the signal strength for the recently inserted record. *R.T* describes the timestamp for the latest record, etc.

The endpoint is a structure for defining callbacks endpoints. It is a named block that is ready to accept asynchronous calls from external sources. For developers, each callback describes an ability to pass the asynchronous request with the given (described) parameters. The "request" here is HTTP GET request.

```
ENDPOINT NEW_RECORD (M, RS)
{
    R(M, RS)
}
```

A Timer is a structure for describing periodically executed code. Timers have names and time interval (in milliseconds):

```
TIMER TM (1000)
{
}
```

A Rule is a logical operator (IF-THEN). Each rule has got a condition and conclusion (block).

```
RULE  R1   R.RSSI< −60
{
}
```

Conditional part includes a regular expression with the above-described relations.

The Modules provide a bridge for the runtime platform. Each module has got a name and a list of output parameters.

```
MODULE COUNTER (count)
```

This description defines a list of output parameters. In other words, it is a list of values that can be used after the method call. All methods correspond to JSP files. The method call is translated into an HTTP request. During a call, we can pass any set of parameters. It is up to the implementation to define how to proceed them. And the method's definition describes output values only. The method call can be direct (immediate) and asynchronous.

The Facts are analogues of statements in procedural languages. In the modern version, LogicIoT supports the following statements:

- Define (add) a record to a relation:

```
R ("38:E7:D8:D3:18:68", −87)
```

- start/stop a timer

```
STOP (TM)
```

- activate/deactivate a rule

```
DEACTIVATE (R1)
```

- check (execute) a rule

```
CHECK(R1)
```

- call a module / asynchronous call a module

```
CALL COUNTER (index, 2)
ACALL COUNTER (name, "test")
```

The Mapping describes a link between the description (declaration) and its implementation. We can describe a mapping for modules and relations.

MAP RELATION R : module1.jsp

MAP MODULE CHECK : module2.jsp

Because our rules form the standard production rule based system, we can use old and well know algorithm like Rete [11] for the processing. A Rete-based expert system builds a network of nodes, where each node (except the root) corresponds to a pattern occurring in the left-hand-side (the condition part) of a rule. The path from the root node to a leaf node defines a complete rule's left-hand-side. Each node has a memory of facts, which satisfy that pattern. This structure presents essentially a generalized tree. As new facts are asserted or modified, they propagate along the network, causing nodes to be annotated when that fact matches that pattern. When a fact or combination of facts causes all of the patterns for a given rule to be satisfied, a leaf node is reached, and the corresponding rule is triggered [28].

In the terms of context-aware programming, our relations play a role of so-called context variables. They can be updated synchronously or asynchronously. The asynchronous update is the prevailing model in IoT applications

Our first implementation is based on custom JSP tags. So, for example, the above-mentioned definition for rule R1 looks that:

```
<iot:Rule name="R1"
          cond="$R.RSSI<=−60">
    some JSP (Java) code
</iot:Rule>
```

The use of Java Server Pages as a foundation lets us easily embed our language modules into web applications. As the next step, the whole JSP file could be wrapped as JSP tags (tag file). It lets us reuse IoT functionality across various projects.

In the connection with the above-described DSL, we should mention, of course, the Business Process Execution Language, commonly known as BPEL or WS-BPEL. It is an XML-based standard markup language that is emerging as the answer to process orchestration requirements [29]. It is a correct question in the context of this paper: why do we choose our own DSL instead of WS-BPEL? For developers, WS-BPEL is a yet another language. Our DSL is finally a set of custom tags. There is no difference between our set of tags and custom tags widely used for web development (e.g., Java Standard Tag Library). Each process here should be organized as a web service. So, on practice, we have to develop proxies for the most of devices. Of course, REST used in our DSL is more lightweight and works faster. Our modules (JSP files) are more flexible than proxies for web services. LogicIoT does not require XML parsing for the each request/response.

LogicIoT contains built-in support for data persistence. In practice, developers very often should use data-oriented extensions of WS_BPEL [30]. For example, authors in [31] describe an extension of IoT. Their WS-BPEL extension includes a when-then construct that process modelers can use to define expected exceptions, using conditions with context variables. They realize the WS-BPEL extension through a language transformation approach. As it adds new activities to process definitions, processes that are executed do not match exactly the process the modeler defined. Nevertheless, the resulting process behaves as expected by the modeler and is independent of the execution engine. By our opinion, it just adds the complexity. We can reach same goals extending the basic programming language.

BPEL (WS-BPEL) is not the only candidate for canceling our own DSL. We should also explain why we do not use the Business Process Model and Notation (BPMN). BPMN is a graphical language for visually defining business processes [32]. BPMN version 2.0 contains enhancements to a graphical notation and meta-model. It presents XML specifications for making such models executable, when properly connected with Web services and/or Java code. In other words, we can directly specify the executable models in BPMN 2.0. Actually, we've followed to the same approach as with BPEL. It is yet another language for developers. Also, there are specific pitfalls for using BPMN with the big amount of asynchronous data sources. The BPMN 2.0 standard does not allow more than one input set and a single Data Input per Service Task [33].

Unfortunately, standards are most likely a long-term solution in complex systems. By our experience, in IoT, they are not answering business needs right now. Unfortunately, the current standardization practice is not suited to accelerating innovation cycles.

IV. CONCLUSION

Data from sensors are becoming widely used by organizations in their business processes. However, to use it, developers have to deal with a massive set of asynchronous processes, associated with the procedure of obtaining data from the individual devices. The approach we present in this paper aims at simplifying the access to IoT information within web applications. Through our DSL, processes can include context data (variables), whose values are updated automatically. The proposed model supports both synchronously and asynchronously updates. And our extension is responsible for the operations required to perform the communication between process instances and sensors.